\documentclass[prl,twocolumn,showpacs]{revtex4}
\usepackage[dvips]{graphicx}
\usepackage{amsmath,amsfonts,amssymb}
\usepackage{dcolumn}
\usepackage{bm}
\usepackage{color}
\usepackage{epsfig}
\usepackage{hyperref}
\begin{document}

\def\jpb{J. Phys. B: At. Mol. Opt. Phys.~}
\def\pra{Phys. Rev. A~}
\def\prb{Phys. Rev. B~}
\def\prl{Phys. Rev. Lett.~}
\def\jmo{J. Mod. Opt.~}
\def\jetp{Sov. Phys. JETP~}
\def\etal{{\em et al.}}

\def\reff#1{(\ref{#1})}

\def\diff{\mathrm{d}}
\def\imagi{\mathrm{i}}

\def\halb{\frac{1}{2}}

\def\beq{\begin{equation}}
\def\eeq{\end{equation}}

\def\energy{{\cal E}}

\def\Ehat{\hat{E}}
\def\Ahat{\hat{A}}

\def\ket#1{\vert #1\rangle}
\def\bra#1{\langle#1\vert}
\def\braket#1#2{\langle #1 \vert #2 \rangle}

\def\vekt#1{\bm{#1}}
\def\vect#1{\vekt{#1}}
\def\vektr{\vekt{r}}

\def\makered#1{{\color{red} #1}}

\def\Im{\,\mathrm{Im}\,}

\def\varphic{\varphi_{\mathrm{c}}}

\def\vxc{v_\mathrm{xc}}
\def\vextop{\hat{v}_\mathrm{ext}}
\def\VC{V_\mathrm{c}}
\def\VHX{V_\mathrm{Hx}}
\def\VHXC{V_\mathrm{Hxc}}
\def\wop{\hat{w}}
\def\Gammaevenodd{\Gamma^\mathrm{even,odd}}
\def\Gammaeven{\Gamma^\mathrm{even}}
\def\Gammaodd{\Gamma^\mathrm{odd}}

\def\Hop{\hat{H}}
\def\HopKS{\hat{H}_\mathrm{KS}}
\def\HKS{H_\mathrm{KS}}
\def\Top{\hat{T}}
\def\TopKS{\hat{T}_\mathrm{KS}}
\def\VopKS{\hat{V}_\mathrm{KS}}
\def\VKS{{V}_\mathrm{KS}}
\def\vKS{{v}_\mathrm{KS}}
\def\Ttildeop{\hat{\tilde{T}}}
\def\Ttilde{{\tilde{T}}}
\def\Vextop{\hat{V}_{\mathrm{ext}}}
\def\Vext{V_{\mathrm{ext}}}
\def\Vopee{\hat{V}_{{ee}}}
\def\psiopdag{\hat{\psi}^{\dagger}}
\def\psiop{\hat{\psi}}
\def\vext{v_{\mathrm{ext}}}
\def\Vee{V_{ee}}
\def\vee{v_{ee}}
\def\nop{\hat{n}}
\def\Uop{\hat{U}}
\def\Wop{\hat{W}}
\def\bop{\hat{b}}
\def\bopdag{\hat{b}^{\dagger}}
\def\qop{\hat{q}}
\def\jop{\hat{j\,}}
\def\vHxc{v_{\mathrm{Hxc}}}
\def\vHx{v_{\mathrm{Hx}}}
\def\vH{v_{\mathrm{H}}}
\def\vc{v_{\mathrm{c}}}
\def\xop{\hat{x}}

\def\varphiexact{\varphi_{\mathrm{exact}}}

\def\fmathbox#1{\fbox{$\displaystyle #1$}}

\title{Rabi Oscillations and Few-Level Approximations\\ in Time-Dependent Density Functional Theory}

\author{M.~Ruggenthaler}
\author{D.~Bauer}
\affiliation{Max-Planck-Institut f\"ur Kernphysik, Postfach 103980, 69029 Heidelberg, Germany}

\date{\today}

\begin{abstract}
The resonant interaction of laser light with atoms is analyzed from the time-dependent density functional theory perspective using a model Helium atom which can be solved exactly. It is found that in exact-exchange approximation the time-dependent dipole shows Rabi-type oscillations of its amplitude. However, the time-dependent density itself is not well described. These seemingly contradictory findings are analyzed. The Rabi-type oscillations are found to be essentially of classical origin. The incompatibility of time-dependent density functional theory with few-level approximations for the description of resonant dynamics is discussed.   
\end{abstract}
\maketitle

Linear response time-dependent density functional theory is widely and successfully applied to calculate absorption spectra of atoms, molecules, and solids \cite{TDDFT}. The predicted transitions between the groundstate and singly excited states are often remarkably accurate \cite{Appel03}, even with simple approximations to the exchange-correlation potential. 
In experiments, a resonant interaction between a laser field and atoms or molecules is routinely used to, e.g., prepare the system in an excited state. The time it takes to transfer the groundstate to an excited state is half a Rabi period ($\pi$-pulse) and can (in simple cases) be calculated using a two-level approximation (TLA, see any text book on quantum optics, e.g., \cite{Scully}).   Time-dependent density functional theory (TDDFT) beyond linear response is, in principle, capable of describing the entire dynamics of the electron density $n(\vektr,t)$ exactly if the exact exchange-correlation potential were known \cite{RG}. With the increasing interest in {\em real-time} quantum dynamics of matter exposed to laser light TDDFT beyond linear response attracts more and more attention  (see \cite{TDDFT} and references therein).

In this Letter we analyze the resonant interaction of laser light with  atoms from the TDDFT perspective. To that end we employ a numerically soluble one-dimensional helium atom as a benchmark model for the corresponding time-dependent Kohn-Sham (TDKS) calculations. We show that the TDKS dipole indeed displays Rabi-type oscillations which, however, are of classical origin, and that the density itself is not properly described in the exact exchange-only approximation. The incompatibility of a TLA with the TDKS equation in the case of resonant interaction is also discussed.

\begin{figure}
\includegraphics[width=0.38\textwidth]{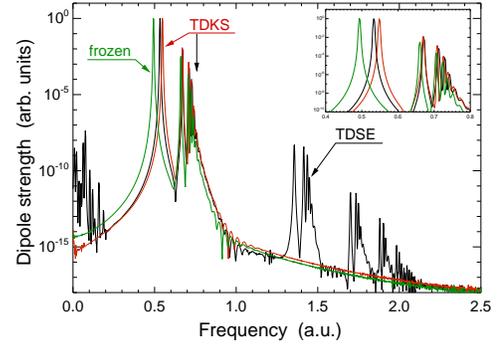} 
\caption{(Color online) Linear response of the He model atom as obtained from the full TDSE (black, labelled 'TDSE'), from the x-only TDKS (red, labelled 'TDKS'), and the frozen x-only KS potential (green, labelled 'frozen'). The vertical arrow indicates the first ionization threshold.  The insert shows a close-up of the transitions to singly excited states. \label{linresp}}
\end{figure}

Let us first introduce the widely used one-dimensional model helium \cite{Model} in which both electrons move along the laser polarization direction only. Softening the Coulomb interaction, $1/\vert\vektr\vert \to 1/\sqrt{1+x^2}$ [atomic units (a.u.) are used], we obtain for the Hamiltonian $H(t)=T+ \Vext(t) + \Vee$ with $ \Vext(t)=-2/\sqrt{1+x^2}  -2/\sqrt{1+{x'}^2} + E(t) [x +x']$, $T=-(1/2) [\partial^2_{x}+ \partial^2_{x'}]$ and $\Vee= 1/\sqrt{1+(x-x')^2}$. Here, the coupling to the laser field is described by the electric field in dipole approximation $E(t)$, $T$ is the kinetic energy, and $\Vee$ the electron-electron interaction. We start from the spin-singlet groundstate. Since there is no spin-dependent external potential the state will remain a spin-singlet state, i.e., $\braket{xx'}{\Psi(t)} = (1/\sqrt{2}) \psi(xx't) [\vert\uparrow\downarrow\rangle - \vert\downarrow\uparrow\rangle]$ with $\psi(xx't)$ being symmetric under the exchange of $x$ and $x'$. The time-evolution is governed by the time-dependent Schr\"odinger equation (TDSE) $\imagi\partial_t\psi(xx't) = H(t) \psi(xx't)$.
The groundstate energy in this system is $\energy_0=-2.238$, the first excited spin-singlet state is at  $\energy_1^{\uparrow\downarrow}=-1.705$. 

The corresponding TDKS equation reads $\imagi\partial_t \varphi(xt) = [-(1/2)\partial^2_{x} + \vKS(xt) ] \varphi(xt)$ with
\beq \label{xonly} \vKS(xt) = -\frac{2}{\sqrt{1+x^2}} +E(t) x  +\int\frac{\vert\varphi(x't)\vert^2\,\diff x'}{\sqrt{1+(x-x')^2}}\eeq
in exact exchange-only approximation, the last term being the Hartree-exchange potential $\vHx$. In this approximation correlation effects are neglected, and the TDKS equation equals the time-dependent Hartree-Fock (TDHF) equation.  
The TDHF wave function is a Slater-determinant, $\braket{xx'}{\Phi(t)} = \varphi(x't) \varphi(xt) [\vert\uparrow\downarrow\rangle - \vert\downarrow\uparrow\rangle]/\sqrt{2}$, and the TDKS density is simply given by $n(xt)=2\vert\varphi(xt)\vert^2$.

The linear response spectra are calculated according to Ref.~\cite{Yabana}. The result is shown in Fig.~\ref{linresp}. The strongest peak is associated with the transition between the groundstate and the first excited singlet-state at $\omega=\energy_1^{\uparrow\downarrow} - \energy_0 = 0.533$, followed by transitions to higher excited states and the first continuum (indicated by the vertical arrow). The TDSE-spectrum also shows transitions to doubly excited states and the corresponding continua. Such transitions are absent in linear response TDDFT employing simple, adiabatic exchange-correlation potentials \cite{Maitra04}. 

The linear response spectrum for the ``frozen'' Kohn-Sham (KS) groundstate potential  $\vKS^{(0)}(x) = -2/{\sqrt{1+x^2}} +\int(n_0(x')\,\diff x')/(2 \sqrt{1+(x-x')^2}) $ (commonly called "bare" KS response) is also included in Fig.~\ref{linresp}, showing the transitions to excited states in  $\vKS^{(0)}(x)$ and illustrating that  $\vKS(xt)-\vKS^{(0)}(x)$ shifts the peaks closer to the correct positions.

\begin{figure}
\includegraphics[width=0.38\textwidth]{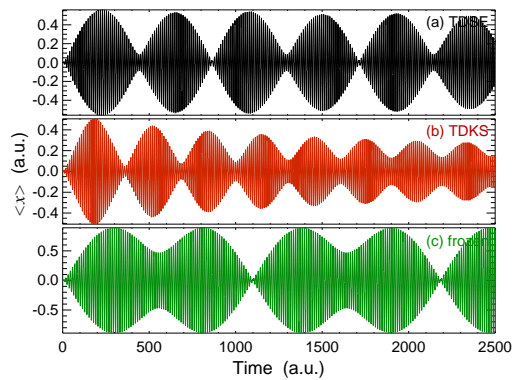} 
\caption{(Color online) Expectation values $\langle x\rangle$ vs time for a resonant excitation with $\Ahat=0.0125$ and (a) $\omega=0.533$ for the TDSE, (b)  $\omega=0.549$ for the TDKS, and (c) $\omega=0.492$ for the frozen KS calculation.  \label{fig:dipoles}}
\end{figure}

Let us now consider a monochromatic laser beam of resonant frequency $\omega=\energy_1^{\uparrow\downarrow} - \energy_0$ and electric field amplitude $\Ehat=\omega\Ahat$. Assuming that the TLA and rotating wave approximation are valid, we expect Rabi oscillations of frequency $\Omega=\Ehat \mu_{10}$ to occur, where 
\[ \mu_{10}= \bra{\Psi_0} (x+x') \ket{\Psi_1} = 2 \int\!\!\!\int\diff x'\diff x\, \psi_0^*(xx') x \psi_1(xx') \]
is the transition dipole matrix element. Its numerical value is $1.1$. The dipole then evolves according
\beq d(t)=\mu_{10} \sin\omega t \sin\Omega t     . \label{eq:dipole} \eeq
Figure~\ref{fig:dipoles} shows $\langle x\rangle (t) = d(t)/2$ as it results from the TDSE, TDKS, and frozen KS calculations. A laser field of vector potential amplitude $\Ahat=0.0125$ was ramped up over two laser cycles and then held constant. The laser frequency was tuned to the resonance $\omega=\energy_1^{\uparrow\downarrow} - \energy_0$, i.e.,  $\omega=0.533$ for the TDSE, $\omega=0.549$ for the TDKS, and $\omega=0.492$ for the frozen KS calculation (all inferred from Fig.~\ref{linresp}). The TDSE result shown in panel (a) displays Rabi oscillations of the envelope of frequency $\Omega=\Ahat \omega \mu_{10}=0.0075$, as expected. At $t=\pi/\Omega\simeq 420$ the excited state is maximally populated and the envelope of the excursion is close to zero. At $t=2\pi/\Omega\simeq 840$ the system is mostly in the groundstate again. A closer inspection of the TDSE result shows that because of ionization and transitions to other states the population of the first excited state after half the Rabi period is only $0.975$ instead of unity. The population of the groundstate after a full Rabi cycle is $0.96$. In the frozen KS calculation [panel (c)] ionization  and the population of other excited states are more pronounced. As a consequence, the excursion envelope does not go to zero at $t=\pi/\Omega$ and the excursion amplitude is overestimated. 
The TDKS calculation in panel (b) shows oscillations of the right amplitude. The Rabi period one infers from the envelope-oscillations is remarkably close to the exact result in (a). For non-resonant driving the amplitude oscillations are absent, as they should.

\begin{figure}
\includegraphics[width=0.4\textwidth]{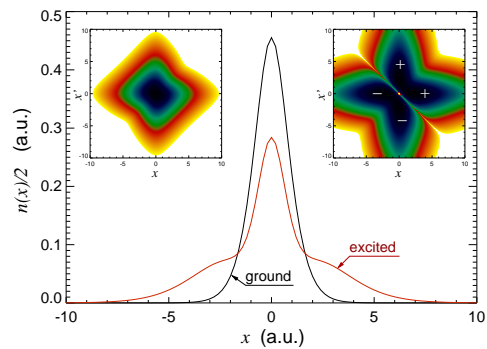} 
\caption{(Color online) Exact groundstate density $n_0(x)/2= \int\diff x'\,\vert \psi_0(xx') \vert^2$ (black, labelled 'ground') and exact excited state density $n_1(x)/2= \int\diff x'\,\vert \psi_1(xx') \vert^2$ (red, labelled 'excited'). The two inserts show contour plots of $\vert \psi_0(xx') \vert^2$ (left) and  $\vert \psi_1(xx') \vert^2$ (right). The signs of $ \psi_1(xx')$ are indicated in the right insert. \label{fig:densities}}
\end{figure}

Figure~\ref{fig:dipoles}b suggests that Rabi oscillations are well described within the TDKS system. If this were true the TDKS density should oscillate between the groundstate density and a density similar to the exact first excited state density shown in Fig.~\ref{fig:densities}. Unfortunately, this is not the case. Examining the TDKS density at time $t\simeq 350$ reveals that it does not assume the shape of the exact excited state density of Fig.~\ref{fig:densities} but rather resembles the initial density again!   Hence, despite an erroneous TDKS density  we observe the Rabi-like oscillations of Fig.~\ref{fig:dipoles}b in its first moment, i.e., in the TDKS dipole.  As for our two-electron system the exact KS orbital corresponding to the excited state density is given by $\varphi_1(x) = \sqrt{n_1(x)/2}$ and thus, according to Fig.~\ref{fig:densities}, has no nodes, the exact KS orbital representing this excited state density must be the {\em groundstate} of a KS potential $\vKS[n_1]$. Hence, even  the exact KS potential will not lead to a population transfer to an {\em excited} KS state but will ``guide''  the density towards the stationary, ``new'' {\em groundstate} density  $n_1$ during a $\pi$-pulse. The exact exchange-only approximation used in (\ref{xonly}) above does not do this. Hence, correlation is needed to describe the density dynamics properly. Whether memory effects \cite{Maitra02} are important in this context will be investigated in a forthcoming paper. 
{Note that the exact KS potential may be simply constructed by inversion for the case of the two-electron spin-singlet system studied here \cite{Lein05}. However, in order to actually {\em identify} memory effects more involved methods are required, e.g., the approach proposed in Ref.~\cite{thiele}.}

\begin{figure}
\includegraphics[width=0.36\textwidth]{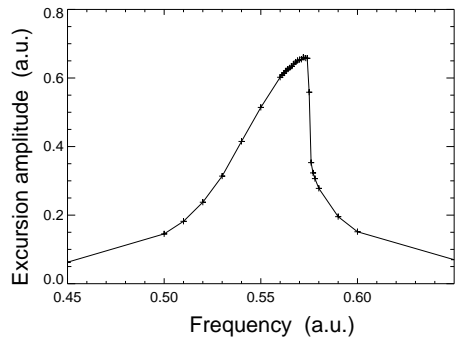} 
\caption{TDKS excursion amplitude $B=\max x$ vs laser frequency for a driver with $\Ahat=0.0125$. Discontinuity and asymmetric peak structure are characteristic of classical anharmonic oscillations \cite{Landau}.
\label{fig:anharmonicpeak}}
\end{figure}

Let us now investigate the origin of the Rabi-like oscillations in the TDKS dipole of Fig.~\ref{fig:dipoles}. Since it is not due to the density-dynamics corresponding to population transfer there must be another explanation. In order to show that the oscillations are classical in 
origin let the motion of the center of mass of the density being approximated
by the dynamics of a point particle in an effective anharmonic potential $v(x) = (\omega^{2}/2) x^{2} + (\alpha/3) x^{3} + (\beta/4) x^{4}$ with $\alpha$ and $\beta$ some constants. The external driver is of the form $E(t)=\Ehat \cos[(\omega+\epsilon) t]$, i.e., $\epsilon$ is the detuning with respect to $\omega$, i.e., the frequency characterizing  the harmonic region of the potential around the origin. The squared excursion amplitude $B^2(\epsilon)=[\max x(t)]^2$ fulfills a cubic equation \cite{Landau}. As soon as a critical driver strength is reached more than one real solution for $B(\epsilon)$ exists, and a discontinuity develops. Figure~\ref{fig:anharmonicpeak} shows that the amplitude of the TDKS dipole as a function of the laser frequency displays exactly this feature. The classical Rabi-like oscillations can intuitively be understood as follows: while being in the harmonic region of the potential the particle is resonantly driven and thus the excursion amplitude increases. However, with increasing excursion amplitude the particle inevitably senses the anharmonicity of the potential. As a consequence the driver is not resonant anymore and the excursion amplitude decreases. Hence the Rabi-like oscillations of the excursion amplitude seen in our TDKS results are essentially of classical origin. Similar oscillations were observed in Josephson junctions \cite{gronbech05}.

Going back to our model helium where the TLA applied to the interacting system, $\ket{\Psi(t)} \simeq a(t) \exp(- \imagi \energy_0 t) \ket{\Psi_{0}} + b(t) \exp(- \imagi \energy_1 t) \ket{\Psi_{1}}$, accurately captures the resonant population transfer to an excited state, the density evolves in time according 
$n(xt)= |a(t)|^{2} n_{0}(x) + |b(t)|^{2} n_{1}(x) + 2 \Re \{a^{*}(t) b(t) \exp(- \imagi \omega t ) \Delta n(x)\}$. Here $\Delta n(x)=\braket{\Psi_{0}|\hat{n}(x)}{\Psi_{1}}$ is real with $\hat{n}(x) = \sum_{\sigma} \hat{\psi}_{\sigma}^{\dagger}(x) \hat{\psi}_{\sigma}(x) $ the density operator in second quantization with spin degrees of freedom $\sigma$. Can we use this accurate, interacting two-level density to construct a  $v_{\mathrm{Hxc}}$ for a corresponding two-level KS scheme?
In order to show that such an approach will fail we start from the fundamental equation for the derivation of the extended Runge-Gross proof (see \cite{vanLeeuwen} or Ch.~2 in \cite{TDDFT}) in one dimension,
\beq
\label{RGproofeqn}
 \partial_{t}^{2} n(xt) = \partial_{x} [ n(xt) \partial_{x} \vext(xt) ]  +q([n];xt), 
\eeq
with $q([n];xt)=\braket{\Psi(t)|\partial_{x}^{2} \hat{T}(x) + \partial_{x} \hat{V}_{\mathrm{ee}}'(x)}{\Psi(t)}$,
the momentum-stress tensor $\hat{T}(x) = \sum_{\sigma}
\{[\partial_{x}\hat{\psi}_{\sigma}^{\dagger}(x)]\partial_{x}\hat{\psi}_{\sigma}(x)-\frac{1}{4}\partial_{x}^{2}[\hat{\psi}_{\sigma}^{\dagger}(x)\hat{\psi}_{\sigma}(x)]\} $ and the interaction term
$\hat{V}_{\mathrm{ee}}'(x)=\sum_{\sigma, \sigma'} \int \diff x'\,
\hat{\psi}_{\sigma}^{\dagger}(x)\hat{\psi}_{\sigma'}^{\dagger}(x')[\partial_{x}
v_{\mathrm{ee}}(|x-x'|)]
\hat{\psi}_{\sigma'}(x')\hat{\psi}_{\sigma}(x)$. With the analogue of (\ref{RGproofeqn}) for the noninteracting system and its state $\ket{\Phi(t)}$ [leading to the same density $n(xt)$], one finds with $v_{\mathrm{KS}}([n]; xt) = v_{\mathrm{ext}}(xt) + v_{\mathrm{Hxc}}([n];xt)$ for the Hartree-exchange-correlation potential
\begin{eqnarray}
\label{HxcPotDef}
 \partial_{x} [ n(xt)\partial_{x} v_{\mathrm{Hxc}}([n]; xt) ] &=&q([n];xt)  
\\
 &&- \braket{\Phi(t)|\partial_{x}^{2} \hat{T}(x)}{\Phi(t)}. \nonumber
\end{eqnarray} 
One could now use the interacting two-level density in \reff{HxcPotDef} to construct $v_{\mathrm{Hxc}}([n]; xt)$.  However, even if one were able to determine $a(t)$ and $b(t)$ from the noninteracting system alone (so that one has not to solve the full interacting problem in the first place) such a TDKS treatment will lead to wrong predictions. Note that with (\ref{RGproofeqn}) and $v_{\mathrm{ext}}(xt) = v_{0}(x) + x  E(t)$ the dipole acceleration of an $N$ particle system reads $\ddot{d}(t) = - N E(t) - \int \diff x \, n(xt) \partial_{x} v_{0}(x)$. Here the term depending on  $\partial_{x} v_{\mathrm{Hxc}}([n]; xt)$  vanishes in accordance with the zero-force theorem \cite{vignale,TDDFT}. With $n_{0}(x)$, $n_{1}(x)$ symmetric and $v_{0}(x)$ even, $\int \diff x \, n_{0}(x) \partial_{x} v_{0}(x) = \int \diff x \, n_{1}(x) \partial_{x} v_{0}(x) = 0$ results, and thus
\begin{eqnarray}
\label{WrongDipoleAcc}
 \ddot{d}(t) + c d(t) = - N E(t),
\end{eqnarray}
where $c=\int \diff x\, (\partial_{x}v_{0}(x))\Delta n (x)/\int \diff x\, x \Delta n(x)$. Equation (\ref{WrongDipoleAcc}) describes a driven harmonic oscillator which [for initial conditions $d(0)=\dot{d}(0)=0$] does not exhibit oscillations of the excursion amplitude as a function of the driver amplitude, i.e., no Rabi-like oscillations. Hence, despite using an accurate density as input not even the dipole is reproduced within such a two-level TDDFT. What went wrong?
It turns out that the introduction of a TLA into (\ref{RGproofeqn}) leads to inconsistencies between Hilbert spaces and Hamiltonians. In order to illustrate this fact it is sufficient to consider the Heisenberg equation of motion for some general, time-independent operator $\hat{O}$,
\begin{eqnarray}
\label{Ehrenfest}
 \partial_{t} \hat{O}_{\mathrm{H}}(t) & = & \imagi [\hat{H}_{\mathrm{H}}(t), \hat{O}_{\mathrm{H}}(t)]  
\\
&= &\imagi \left( \hat{U}^{-1}(t) [\hat{H}(t), \hat{O}]\hat{U}(t)  \right), \nonumber
\end{eqnarray}
with $\hat{O}_{\mathrm{H}}(t) = \hat{U}^{-1}(t) \hat{O}\hat{U}(t)$ and $\hat{U}(t) = \mathcal{T}\left(\exp(-\imagi \int_{0}^{t} \diff t'\, \hat{H}(t')) \right)$ the time evolution operator. Making use of (\ref{RGproofeqn}) and introducing a TLA amounts to calculating the commutators in expressions like (\ref{Ehrenfest}) in the  full Hilbert space while the time evolution is performed in a reduced Hilbert space, i.e., in our case in a two-level subspace,
\begin{eqnarray}
\label{2LEhrenfest}
&&\imagi \left( \hat{U}^{-1}_{\mathrm{2}}(t) [\hat{H}(t), \hat{O}]\hat{U}_{\mathrm{2}}(t)  \right) 
\\
&&= \imagi \left\{ \hat{U}^{-1}_{\mathrm{2}}(t)\hat{\bm{1}}_{2}\left(\hat{H}(t) \hat{\bm{1}} \hat{O} - \hat{O}\hat{\bm{1}} \hat{H}(t) \right)\hat{\bm{1}}_{2}\hat{U}_{\mathrm{2}}(t) \right\}, \nonumber
\end{eqnarray}
with $\hat{\bm{1}} = \sum_{k=0}^{\infty} \ket{\Psi_k}  \bra{\Psi_k}$, $\hat{\bm{1}}_{2} = \sum_{k=0}^{1} \ket{\Psi_k}  \bra{\Psi_k}$ and $\hat{U}_{2}(t) = \mathcal{T}\left(\exp(-\imagi \int_{0}^{t}\diff t'\, \hat{\bm{1}}_{2}\hat{H}(t')\hat{\bm{1}}_{2}) \right)$.
Instead, for a consistent TLA 
\begin{eqnarray}
 &&\partial_{t} \left( \hat{U}_{2}^{-1}(t) \hat{O} \hat{U}_{2}(t) \right) 
\\
&&= \imagi \left\{ \hat{U}_{2}^{-1}(t) \hat{\bm{1}}_{2}\left( \hat{H}(t) \hat{\bm{1}}_{2} \hat{O} - \hat{O} \hat{\bm{1}}_{2} \hat{H}(t)   \right)\hat{\bm{1}}_{2}\hat{U}_{2}(t)  \right\} \nonumber
\end{eqnarray}
holds, with a different commutator leading to a different equation of motion. In particular, the equation for $\partial^2_t n(xt)$ is different from \reff{RGproofeqn} if the Hamiltonian is restricted to a two-level subspace \cite{Berman73}. Obviously, our analysis not only applies to a TLA but to any {\em finite}-level approximation.

{So far we tried to construct $v_{\mathrm{Hxc}}$ by applying a TLA to Eq.~\reff{HxcPotDef}. On the other hand, by inversion \cite{Lein05} we can determine the KS potential $v_{\mathrm{KS}}$ generating {\em exactly} a {\em given} two-level density $n(xt)$.}  The associated dipole acceleration then reads
$\ddot{d}(t) = - \int \diff x\, n(xt) \partial_{x} v_{\mathrm{KS}}([n];xt)$. Subtracting from the KS potential the physical external potential defines $v_{\mathrm{Hxc}}$, and we obtain $\ddot{d}(t)+ \int \diff x \, n(xt)\partial_{x} v_{0}(x) + \int \diff x \, n(xt)
\partial_{x} v_{\mathrm{Hxc}}([n];xt)  = \ddot{d}(t)+ c d(t) + \int \diff x \, n(xt)
\partial_{x} v_{\mathrm{Hxc}}([n];xt)  =  -NE(t)$. By construction, the two-level density leads to the correct two-level dipole acceleration. This can only be possible if the term depending on $v_{\mathrm{Hxc}}$ contributes. Otherwise the same problem as with Eq.~(\ref{WrongDipoleAcc}) discussed above arises. However, a nonvanishing contribution from $\int \diff x \, n(xt)
\partial_{x} v_{\mathrm{Hxc}}([n];xt)$ is only possible if $v_{\mathrm{Hxc}}$ does not describe internal forces only and thus violates the zero-force theorem \cite{vignale,TDDFT}. As a consequence, the external potentials of the interacting and the noninteracting system cannot be kept equal. In fact, there is no local external potential $v_\mathrm{ext}(x)$ that supports just two levels. {Nevertheless, a $v_{\mathrm{Hxc}}$, which violates the zero-force theorem, may be acceptable as an approximation.}

In conclusion, we investigated resonant dynamics in the exact-exchange approximation for the simple but numerically exactly solvable case of a one-dimensional model helium atom. Although the dipole shows Rabi-type oscillations the density-dynamics of the population transfer process is not properly described. As a consequence, the dipole spectra calculated using exact exchange-only TDDFT may be sufficiently accurate while the real-time dynamics of the density is erroneous. The incompatibility of few-level approximations with TDDFT to describe resonant density-dynamics was analyzed. Since there is an increasing interest in strongly driven real-time quantum dynamics of matter the development of exchange-correlation potentials capable of describing resonant charge transfer is particularly important and will be the subject of future work.

This work was supported by the Deutsche Forschungsgemeinschaft.

\end{document}